\documentclass[a4paper]{jpconf}
\usepackage{graphicx}
\usepackage{amsfonts}
\usepackage{url}
\begin{document}
\title{Improved selective background Monte Carlo simulation at Belle II with graph attention networks and weighted events}

\author{B Yu, N Hartmann, L Schinnerl and T Kuhr}

\address{Ludwig Maximilian University of Munich, Geschwister-Scholl-Platz 1, 80539 Munich, Germany}

\ead{Boyang.Yu@physik.uni-muenchen.de}

\begin{abstract}
When measuring rare processes at Belle II, a huge luminosity is required, which means a large number of simulations are necessary to determine signal efficiencies and background contributions. However, this process demands high computation costs while most of the simulated data, in particular in case of
background, are discarded by the event selection. Thus, filters using graph neural networks are introduced at an early stage to save the resources for the detector simulation and reconstruction of events discarded at analysis level. In our work, we improved the performance of the filters using graph attention and investigated statistical methods including sampling and reweighting to deal with the biases introduced by the filtering.
\end{abstract}

\section{Introduction}
The standard model is to date the most thoroughly tested fundamental theory that describes nature. Nonetheless, there are still many primary questions to be answered. To complement the standard model, many new physics scenarios have been proposed and high energy physics experiments have been designed to search for new particles and new processes. To enhance the chances of detecting interesting events with low cross sections, the SuperKEKB accelerator~\cite{kek} is anticipated to achieve a high luminosity, with the Belle II experiment~\cite{abe2010belle} producing a large volume of data for analysis. To extract physical information, Monte Carlo simulations of even larger sizes are compared with the real data. 

\begin{figure}[h]
    \centering
    \includegraphics[scale=0.4]{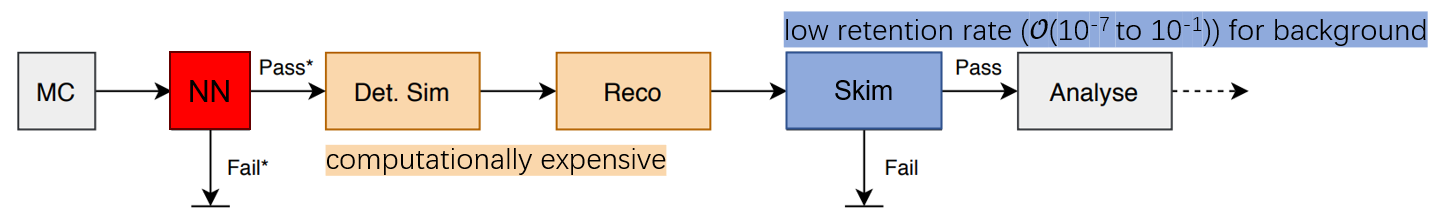}
    \caption{The use of NN filter in the data flow.}
    \label{dataflow}
\end{figure}

Although the simulations are computationally expensive, most of the simulated events are eventually discarded  due to the selection requirements necessary for the data analysis, called skimming (Figure~\ref{dataflow}). Therefore, it has been proposed by James Kahn~\cite{james} to use a filtering neural network to pre-select events that have the potential to pass the skim. In the first part of this study, we focused on the efficacy of attention mechanisms to improve the performance of the filter.

The pre-selection process will however introduce biases due to wrong discarding of events. To address this issue, a contribution of distance correlation to the loss term was presented to punish biases during training~\cite{Yannick}. In this work, we investigated two further methods for dealing with this problem.

\section{Dataset preparation}
Following the previous studies mentioned in the introduction, we selected a dataset based on the Full Event Interpretation (FEI)~\cite{FEI} skim from the Belle II experiment. The FEI skim was chosen as the study object mainly because of its low, but sufficient retention rate (around 6\%), providing enough data to train classification models. The selected skim selection is used by various studies, meaning that the achievement of these works can be directly used by others.

The dataset is constructed from simulated $\Upsilon(4S)\to B^0\bar{B}^0$ events where the skim selection corresponds to the hadronic $B^0$ FEI. Each instance is labeled with a \textit{True} or \textit{False} value, indicating whether it is able to pass the skim or not. For this study, a training set of $\mathrm{9\times10^5}$ samples is used, and $\mathrm{1\times10^5}$ and $\mathrm{5\times10^5}$ samples are reserved for the validation set and test set, respectively. For each dataset there is an equal amount of \textit{True} and \textit{False} labeled instances. The training and validation sets are used for the learning process of the neural network models and gradient boosting decision tree (GBDT)~\cite{gbdt} classifiers, while the test set is only used for the final evaluation of different performance metrics, including classification accuracy, speedup, and feature importance.

The information generated by Monte Carlo has a rooted tree structure in which each node represents a single particle and the connection between particles is indicated by the mother particle index on each node. For the building of data sets, a number of generator-level features of each particle are selected, which can be divided into two categories based on their usage: \textit{Generated Variables} and \textit{Physics Observables}. 

The \textit{Generated Variables} play a crucial role in constructing the input for the graph neural network structures, by providing the information to distinguish between background and signal decays. These features will be collected over nodes from graphs by the neural network to calculate global features that are decisive in the final classification. The generated variables include \textbf{PDG ID}~\cite{pdg}, \textbf{Mother array index}, \textbf{Production Time}, \textbf{Energy}, \textbf{Momentum}, and \textbf{Position}.

The \textit{Physics Observables} on the other hand are not used in the training of the neural networks, since they are not available after Monte Carlo event generation. These features are important for physics analyses and it is therefore vital to study any potential biases of their distributions. Based on the previous study~\cite{Yannick}, a set of 14 physics observables that exhibit the strongest biases among 29 features defined in~\cite{james} were selected, among them is the beam-constrained mass $\mathrm{\textbf{M}_{bc}}$, which is a crucial variable for the FEI skim.

\begin{figure}[h!]
    \centering
    \includegraphics[scale=0.3]{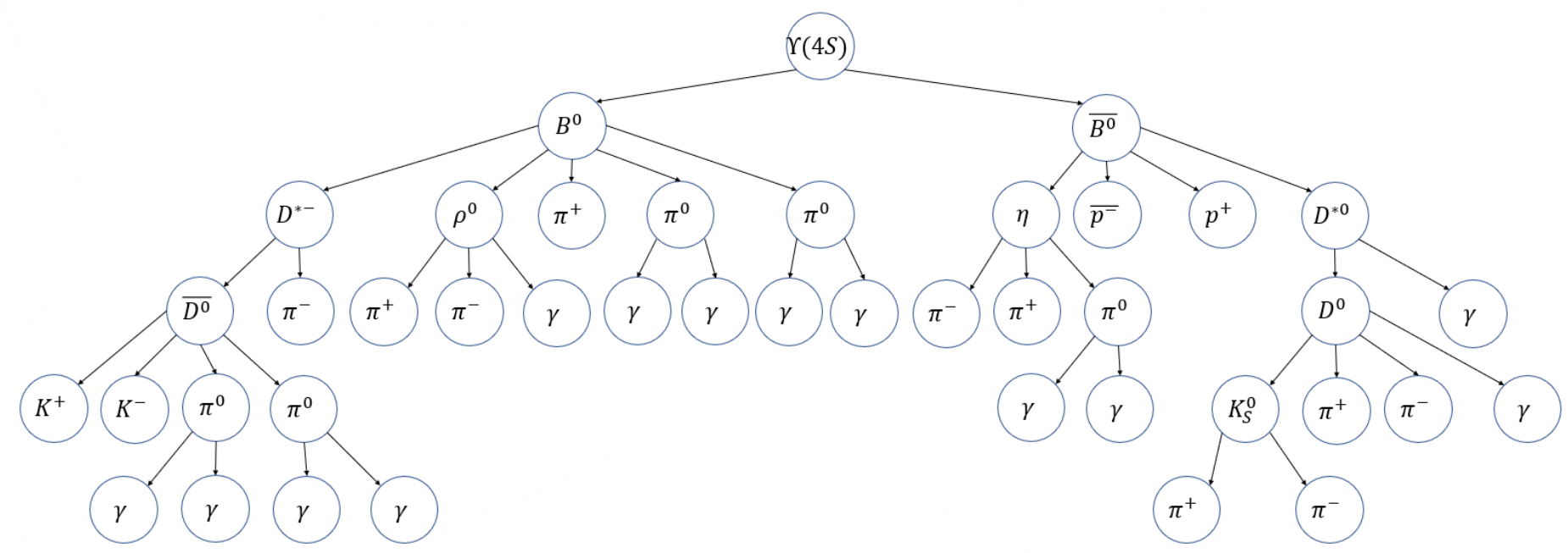}
    \caption{An example of $\Upsilon(4S)\to B^0\bar{B}^0$ decay}
    \label{decay}
\end{figure}

Decay events are represented as graphs by utilizing the mother particle array index carried by each particle, with the expectation that all decays are generated by $\Upsilon(4S)\to B^0\bar{B}^0$ (as shown in Figure~\ref{decay}). The node features are the \textit{Generated Variables}. One decay event can be fully represented by a single graph, with a boolean label indicating whether the decay satisfies the FEI skim.

\section{Neural network filter}
\begin{figure}[htbp]
    \centering
    \begin{minipage}[t]{0.46\textwidth}
        \centering
        \includegraphics[width=7cm]{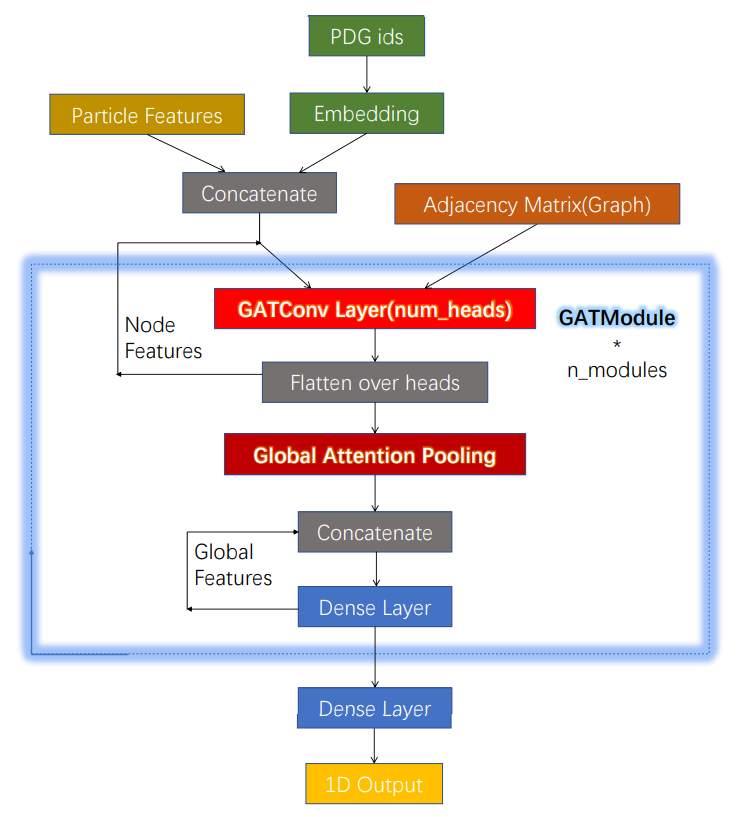}
    \end{minipage}
    \begin{minipage}[t]{0.46\textwidth}
        \centering
        \includegraphics[width=7cm]{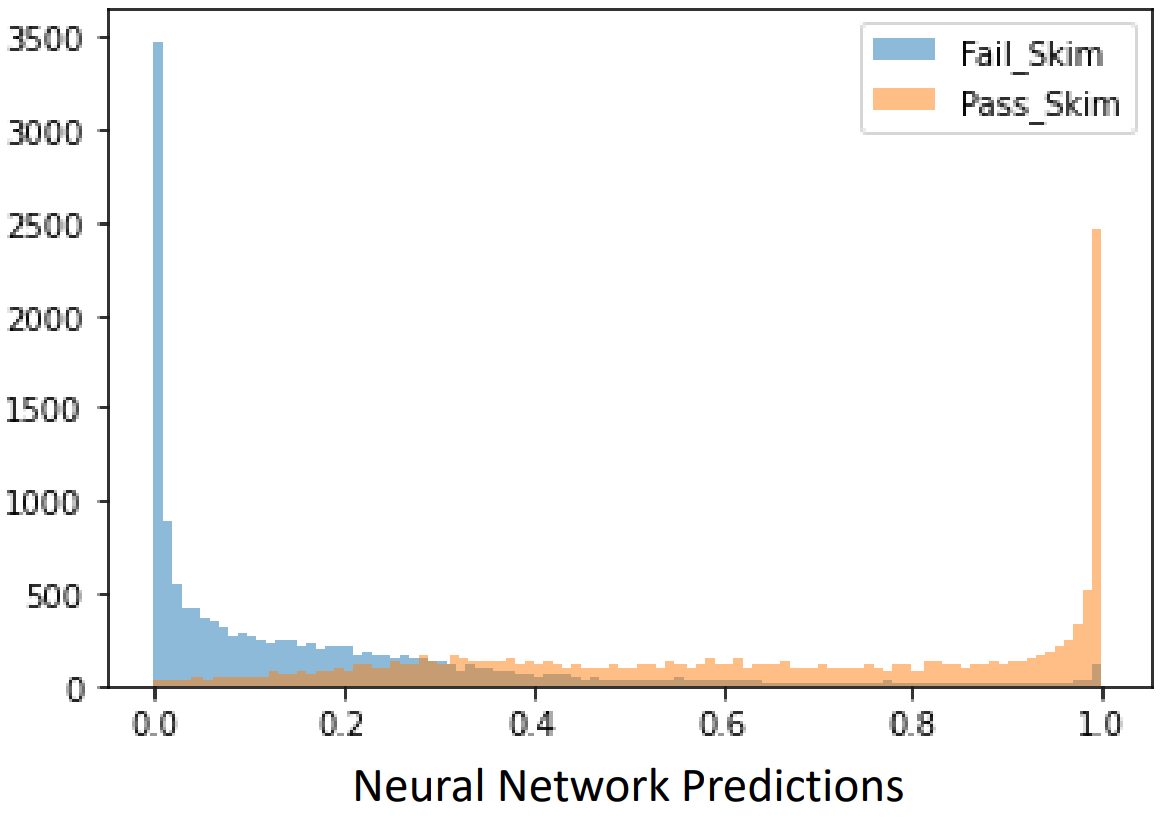}
    \end{minipage}
    \caption{(left) Final structure of the neural network filter. (right) Performance.}
    \label{gat}
\end{figure}

Graph Convolutional Networks (GCN)~\cite{gcn} used in the previous study~\cite{Yannick} showed strong ability to perform the classification. An advanced version of GCN benefiting from attention mechanisms~\cite{attention} was proposed known as Graph Attention Networks (GAT)~\cite{gat}. Through experiments, we demonstrated that the GAT architecture (Figure~\ref{gat} left) performs better than its predecessor. 

We compared different construction approaches and found an optimal version. In this approach, the PDG IDs of particles are first fed through an embedding layer to convert them to tokens represented in a 540-dimensional space. These tokenized IDs are then concatenated with the other generated variables to form the initial node features. The preprocessed node features are then fed into the GAT module. In addition to the node features, the GAT module takes the adjacency matrix representing the connections between the particles in the graph as input. The adjacency matrix remains fixed during the training and provides the GAT module with information on how particles are connected in each graph.

The GAT module is a crucial component for updating both node features and global features. After the Graph Attention (GATConv Layer in the figure), the node features on each node are updated and used as the new input node features for the next layer. Additionally, the updated node features are fed into the Global Attention Pooling layer, which sums the features from all nodes in the graph with inferred weights and generates a set of graph structure-independent global features. These global features are then concatenated with the global feature output from the previous layer, which is initialized to 0, and mapped to a fixed number of dimensions by a Dense layer. This process results in a new set of global features that are used to update the next layer. Finally, after passing the last layer, the global features are mapped to a single neural network output value, the score $p\in[0,1]$. 

The search for the best model and optimal hyperparameters was carried out based on several criteria, including the accuracy on validation set, loss on both validation and training sets, training time on the training set, and ROC curve and AUC values on the test set. The best AUC value was improved from 0.9083 to 0.9122.

\section{Bias reduction}
\begin{figure}[htbp]
    \centering
    \begin{minipage}[t]{0.4\textwidth}
        \centering
        \includegraphics[width=6.4cm]{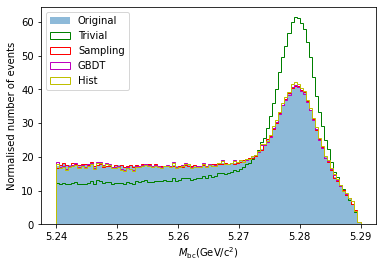}
    \end{minipage}
    \begin{minipage}[t]{0.4\textwidth}
        \centering
        \includegraphics[width=6.4cm]{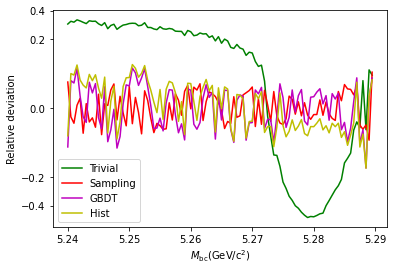}
    \end{minipage}
    \caption{Reproduced $M_\mathrm{bc}$ with trivial selection (green), sampling (red), GBDT reweighting (purple) and Histogram reweighting (yellow).}
    \label{compare}
\end{figure}

After the neural network filter, each event is assigned a score indicating the probability of this event passing the following skims. The trivial approach is to define a threshold, where the events with a higher score are kept. However, this selection will reject some samples that have been able to pass the skims, resulting in false negative events and leading to deviations, or biases (Figure~\ref{compare} green), in the following analysis. The objective of this part is to identify and minimize biases by utilizing statistical tools.

The first step is to measure the biases with the help of the 14 physics observables attached to each event. The histograms of each variable of all the \textit{True} labeled events are compared with those of all the \textit{True-Positive} events, whose scores are required to be larger than the threshold. Figure~\ref{compare} (red) shows the deviation of the reproduced $M_\mathrm{bc}$ before any post-processing. 

One approach of mitigating the biases is to use a sampling method, where events are randomly selected, and weighted by the inverse sampling probability, and no threshold is needed. However, this leads to a lower effective sample size due to higher statistical uncertainty when using weighted events and must be taken into account to estimate the performance gain. To train the neural network filter with this in mind, the loss function is chosen to be the speedup metric, which represents the time reduction of producing the same effective sample size of events. Maximizing the speedup function during training optimizes the network to assign the best sampling weight to each event. The agreement between True-Pass and sampled distributions is satisfactory, as shown in Figure~\ref{compare} (red), using $M_\mathrm{bc}$ as an example. Although the network is trained with speedup as the loss function, the final speedup achieved is only 2.1, indicating that only half of the production time can be saved for each event without introducing any bias.

To achieve a higher speedup, the reweighting method was studied. In this approach, the neural network filter is trained to predict the probability of an event to pass. Filtered events are selected by a fixed threshold on this output probability. Then, a GBDT classifier is trained to predict the probability of \textit{True-Positive} events in the set of all events with \textit{True} label. The weights, or the inverse conditional probability of \textit{True-Positives}, are derived from the GBDT output using two methods. For {\it GBDT Reweighting} the inverse of the classifier output is directly used as the weight, while for {\it Histogram Reweighting} the weight is derived from histograms of the classifier output. The performance of both methods is shown in Figure~\ref{compare} (purple and yellow). The bias of $M_\mathrm{bc}$ is similar to the one using sampling method since it is one of the variables used in the training of the GBDT classifier. For some other variables the biases are higher.

The performances of all the methods were compared according to their highest achieved speedup and their highest biases in the reconstructed physics observables, which are measured by Kolmogorov–Smirnov statistic (KS Test). A trade off between the speedup and the bias minimizing is presented in the table ~\ref{concl}.

\begin{table}[h!]
    \centering
        \begin{tabular}{|c|c|c|}
        \hline
        Method & Best Speedup & Bias (KS Test) \\
        \hline
        Sampling & 2.1 & $\sim0$ \\
        \hline
        GBDT Reweighting & 5.5 & $\sim8\sigma$ \\
        \hline
        Histogram Reweighting & 6.5 & $\sim15\sigma$ \\
        \hline
    \end{tabular}
    \caption{Conclusion of the different weighting methods.}
    \label{concl}
\end{table}

\section{Summary}
We improved the graph neural network filter by incorporating attention mechanism and synchronising the update of node and global features. To avoid the bias generated by discarding \textit{False-Negative} events, we introduced importance sampling, GBDT reweighting and Histogram reweighting, based on statistical uncertainty, to compensate the selected events, either randomly or through a threshold, with higher weights. The sampling method has the strongest bias-mitigation ability but achieves a speedup factor of two. The reweighting methods offer higher speedup factors of around 6 but still have small bias and require careful tuning and validation.

\section{Acknowledgments}
This work was supported by the German Federal Ministry of Education and Research (BMBF) in the collaborative project IDT-UM (Innovative Digitale Technologien zur Erforschung von Universum und Materie).

\end{document}